\documentclass[twocolumn,showpacs,preprintnumbers,amsmath,amssymb]{revtex4}

\newcommand {\be}{\begin{equation}}
\newcommand {\ee}{\end{equation}}
\newcommand {\bey}{\begin{eqnarray}}
\newcommand {\eey}{\end{eqnarray}}
\usepackage{epsfig}
\usepackage{amsfonts}
\usepackage{amsmath}
\usepackage{mathbbol}

\begin{document}

\title{Measurement contextuality is implied by macroscopic realism}

\author{Zeqian Chen}
\affiliation{Wuhan Institute of Physics and Mathematics, Chinese Academy of Sciences, 
30 West District, Xiao-Hong-Shan, Wuhan 430071, China}

\author{A. Montina}
\affiliation{Perimeter Institute for Theoretical Physics, 31 Caroline Street North, 
Waterloo, Ontario N2L 2Y5, Canada}

\date{\today}

\begin{abstract}
Ontological theories of quantum mechanics provide a realistic description of single 
systems by means of well-defined quantities conditioning the measurement outcomes. 
In order to be complete, they should also fulfil the minimal condition of macroscopic 
realism. Under the assumption of outcome determinism and for Hilbert space dimension 
greater than two, they were all proved to be contextual for projective measurements. 
In recent years a generalized concept of non-contextuality was introduced that 
applies also to the case of outcome indeterminism and unsharp measurements. It was 
pointed out that the Beltrametti-Bugajski model is an example of measurement 
non-contextual indeterminist theory. Here we provide a simple proof that this model is 
the only one with such a feature for projective measurements and Hilbert space dimension 
greater than two.  In other words, there is no extension of quantum theory providing more 
accurate predictions of outcomes and simultaneously preserving the minimal labelling of 
events through projective operators. As a corollary, non-contextuality for projective 
measurements implies non-contextuality for unsharp measurements. 
By noting that the condition of macroscopic realism requires an extension of
quantum theory, unless a breaking of unitarity is invoked, we arrive at the conclusion 
that the only way to solve the measurement problem in the framework of an ontological 
theory is relaxing the hypothesis of measurement non-contextuality in its generalized 
sense.
\end{abstract}


\maketitle

\section{Introduction}

Quantum mechanics provides an operationally complete and minimal description of states, 
transformations and events. Procedures that cannot be statistically discriminated 
are described in the quantum formalism by the same mathematical objects. This is
for example the case of the quantum state, which contains only the statistically
significant information about the preparation protocol. The same holds for events
of a von Neumann measurement, which are labelled by projective operators $\hat E_k$.
This description
is operationally complete since the probability $p(\hat E_k)$ of an event
$\hat E_k$ does not depend on any supplementary information, such as the 
complete set of events $\{\hat E_1,\hat E_2,...\}$ being measured. 
It is also minimal because events labelled with different projectors have different 
statistical weights for some preparation state.
For reasons of economy, one could desire to preserve the same minimal
description also in ontological theories of quantum mechanics,
traditionally known as hidden variable theories.

Originally, these theories were designed for the purpose of introducing determinism and 
realism in quantum mechanics~\cite{bellSp}. Whereas in the standard interpretation the 
state represents the overall statistically significant information about the preparation 
procedure, an ontological theory provides a realistic description of the actual 
state of affairs of a single system. Employing the recent terminology~\cite{spekkens0}, 
we call these states {\it ontic states} in order to distinguish them from the quantum 
states. In the framework of an ontological theory they condition the probabilities of 
events. The theory is said {\it outcome deterministic} if the ontic state determines with 
certainty the result of a measurement, that is, if the conditional probabilities of events are 
zero or one. For example this is the case of the de Broglie-Bohm theory, where the ontic state 
is identified with the quantum mechanical wave-function and additional variables describing
the actual positions of the particles. 

Additionally, an ontological theory
should also satisfy the minimal condition of {\it macroscopic realism}~\cite{leggett}
in order to be complete, that is, if there are two or more macroscopically distinct 
states available to a macroscopic system, then the ontological theory must attribute 
one of these states to the system. In general a theory with outcome determinism, such as 
the de Broglie-Bohm theory, fulfils this criterion. The Beltrametti-Bugajski 
model~\cite{BB} is a counter-example of ontological model for measurements that does 
not provide a macroscopic realist description, unless the postulate of wave-function
collapse of the standard interpretation is retained in the model.

Since in the quantum formalism an event is labelled by the projective operator
$\hat E_k$, it seems natural to keep this structure also in the ontological 
theory. However the Kochen-Specker theorem establishes that this is impossible 
for Hilbert space dimensions greater than $2$ and outcome determinism~\cite{KS}. 
The occurrence of the event $\hat E_k$ for a given ontic state depends in general 
on the whole set of projectors involved in the measurement. This feature is called 
{\it contextuality}. 
In recent years a generalized concept of non-contextuality was 
introduced that allows for outcome indeterminism and apply to
unsharp measurements~\cite{spekkens}, that it, positive operator
valued measurements (POVM).
Furthermore this definition applies also to states
and transformations. Essentially, an ontological theory is
non-contextual for states, transformations and measurement
events if the probability distributions, the transition rates and
the conditional probabilities associated respectively with states,
transformations and events preserve the same minimal labelling
of quantum mechanics.
In Ref.~\cite{spekkens} it was shown that non-contextuality for
state preparation is incompatible with quantum statistics. This
can be seen also as a consequence of the fact that the probability 
distribution associated with a pure quantum state $|\psi\rangle$ 
cannot be a quadratic function of $|\psi\rangle$~\cite{montina}. 
However
there is not a similar constraint for measurements~\cite{spekkens}. 
Indeed the Beltrametti-Bugajski model~\cite{BB} is a simple 
example of indeterminist theory that is measurement non-contextual.

In this article we report a simple proof that the Beltrametti-Bugajski
model is essentially the only theory that is non-contextual for projective 
measurements. In other words, we prove that no extension of quantum
theory exists providing more accurate predictions of outcomes
and simultaneously preserving the minimal labelling of events
through projective operators. These findings are closely related 
to a recent result reported by Colbeck and Renner~\cite{colbeck0}. 
They conclude that every extension of quantum theory with improved
predictions of outcomes is incompatible with the hypothesis that
the measurement parameters in a region of spacetime, $\Omega$, can be 
chosen to be statistically independent of any variable that does not lie
in the future lightcone of $\Omega$. This hypothesis implies a particular
condition of non-contextuality for measurements.

As a corollary of our result, we show that non-contextuality for unsharp 
measurements is implied by non-contextuality for sharp measurements (the 
projective ones). Finally, we arrive at the conclusion that the only way to 
solve the measurement problem in the framework of an ontological theory and 
fulfil the minimal condition of {\it macroscopic realism} without breaking
the unitarity of evolutions is relaxing the 
generalized condition of measurement non-contextuality. 
In Sec.~\ref{sec1} we introduce the general properties that an ontological 
theory has to satisfy and prove the theorem of uniqueness considering only 
projective measurements. In Sec.~\ref{sec2} the discussion is extended to 
the case of positive operator valued measurements. In Sec.~\ref{sec3} we 
discuss the results and their relation to the question tackled in 
Ref.~\cite{colbeck0}. First, we show that
any extension of quantum theory cannot be accomplished by preserving
measurement non-contextuality. Then, by noting that such an extension is 
necessary in order to introduce macroscopic realism, we infer that 
measurement contextuality is implied by macroscopic realism. 
The conclusions are drawn in the last section.
The main message of this paper is showing that {\it macroscopic realism}
can turn to be a very useful ingredient for establishing general
properties that an ontological theory must satisfy.

\section{Non-contextuality for projective measurements}
\label{sec1}
In an ontological theory a quantum state $|\psi\rangle$ is associated 
with a probability distribution on an ontological space whose elements
are denoted by $X$. In general,
the probability distribution could depend on the context of the 
preparation~\cite{spekkens}, thus we label it with an additional 
variable $\eta$ and define the map
\be
|\psi\rangle\rightarrow \{\rho(X|\psi,\eta)\}
\ee
that associates quantum states with sets of distributions. If there 
is no dependence on $\eta$, the ontological model is said to be 
{\it non-contextual} for preparation of pure states.
Later on it will be shown that this kind of non-contextuality
is implied by measurement non-contextuality.
The probability distributions satisfy the conditions
\bey\label{prop1}
\rho(X|\psi,\eta)\ge0, \\
\label{prop2}
\int dX\rho(X|\psi,\eta)=1.
\eey

In the quantum formalism, a measurement is associated with a
set of commuting projectors $\{\hat E_1,\hat E_2,...\}$ representing
the events. Each complete set of events satisfies the relation
\be
\sum_k \hat E_k=\hat{\mathbb{1}}.
\ee

In the ontological model the probability $P(\hat E_k|X)$ of an
event $\hat E_k$ is conditioned by the ontic state $X$. It
satisfies the inequalities
\be\label{prop3}
0 \leq P (\hat E_k| X) \leq 1
\ee
and, for each complete set $\{\hat E_k\}$ of commuting projectors,
the identity
\be\label{prop4}
\sum_k P(\hat E_k|X)=1.
\ee
We have explicitly employed the hypothesis of measurement
non-contextuality by assuming that the probability of an event
$\hat E_k$ does not depend on the whole set of projective operators.
The projector $\hat E_1$ can for example be an element of the
set $\{\hat E_1,\hat E_2,\hat E_3,...\}$ or 
$\{\hat E_1,\hat E_2',\hat E_3',...\}$, but the conditional
probability does not depend on this change of context. 
Gleason's theorem states that a probability distribution satisfying 
properties~(\ref{prop3}-\ref{prop4}) has the form 
\be\label{gleason0}
P(\hat E_k|X)=\mathrm{Tr}[\hat E_k\hat\rho(X)]
\ee
for some Hermitian operator $\hat\rho(X)$, provided that the
Hilbert space has dimension $3$ or greater~\cite{G}.

The ontological theory is equivalent to quantum mechanics if
\be\label{prop5}
\int dX P(\hat E_k|X)\rho(X|\psi,\eta)=\langle\psi|\hat E_k|\psi\rangle.
\ee

The Beltrametti-Bugajski model trivially satisfies 
properties~(\ref{prop1}-\ref{prop5}). The space of ontic states is the
projective Hilbert space and the probability distribution associated with a 
quantum state $|\psi\rangle$ is
\be\label{state_BB}
\rho(X|\psi) =\delta(X-\psi),
\ee
$\psi$ being the ray of $|\psi\rangle$.
The conditional probability for an event $\hat E_k$ given the ontic
state $X$ is
\be
P(\hat E_k|X)=\langle X|\hat E_k|X\rangle.
\ee

It is simple to prove that any ontological theory that is non-contextual
for projective measurement is essentially equivalent to the 
Beltrametti-Bugajski model. 
\newline
{\bf Theorem 1.} Given a quantum system with a Hilbert space dimension
greater than $2$, any associated ontological theory that is non-contextual
for projective measurements is equivalent to the Beltrametti-Bugajski theory
after a suitable coarse graining of the ontological space.

Proof: Since the ontological theory is non-contextual, then 
any complete set of projective measurements is associated with
conditional probabilities satisfying properties~(\ref{prop3}-\ref{prop4}). 
Gleason's theorem~\cite{G} implies
that there exists a trace-one Hermitian operator $\hat\rho(X)$ such that
\be\label{gleason}
P(\hat E_k|X)=\mathrm{Tr}[\hat E_k\hat\rho(X)],
\ee
for any set $\{\hat E_k\}$ of projective measurements. Let us consider 
the event $\hat E_\psi=|\psi\rangle\langle\psi|$.
If $X$ is in the support of $\rho(X|\psi,\eta)$ for some context $\eta$, 
then the conditional probability $P(\hat E_\psi|X)$ is equal to $1$.
Indeed we have from Eq.~(\ref{prop5}) that
\be\label{PP}
\int dX P(\hat E_\psi|X)\rho(X|\psi,\eta)=1
\ee
Because of properties~(\ref{prop1},\ref{prop2},\ref{prop3}) and Eq.~(\ref{PP}),
\be\label{impl1}
\rho(X|\psi,\eta)\ne0 \Rightarrow P(\hat E_\psi|X)=1.
\ee
This implication is intuitively obvious, if a system is prepared
in a quantum state $|\psi\rangle$ and its ontic state is $X$,
then the probability of obtaining the state $|\psi\rangle$ given
$X$ is $1$.
Thus, since $\hat\rho(X)$ is a positive trace-one Hermitian operator, 
we have from Eqs.~(\ref{gleason},\ref{impl1}) that
\be\label{map_os_qs}
\rho(X|\psi,\eta)\ne0\Rightarrow\hat\rho(X)=|\psi\rangle\langle\psi|.
\ee
Equation~(\ref{map_os_qs}) says that if $X$ is in the support
of the distribution $\rho(X|\psi,\eta)$
then the Hermitian operator $\hat\rho(X)$ is equal to the
pure density operator $|\psi\rangle\langle\psi|$. As a consequence, 
the support of two probability distributions associated with different 
rays are not overlapping. Furthermore, the conditional 
probabilities, given by Eq.~(\ref{gleason}), are constant on the support
of each probability distribution $\rho(X|\psi,\eta)$, that is,
they are not sensitive to the fine structure of the distributions
inside the support.
This allows us to perform a coarse graining of the ontological space, 
dividing it
in equivalent classes $\tilde X$ and associating each class with a 
ray $\psi$,
\be
\tilde X\leftrightarrow \psi.
\ee
The state $\tilde X$ is the union of the supports of the distributions
$\rho(X|\psi,\eta)$ with $\psi$ fixed and $\eta$ spanning every possible 
value. The coarse grained probability distribution on $\tilde X$ is
\be\label{prob_BB}
\rho(\tilde X|\psi)=\delta(\tilde X-\psi)
\ee
and does not depend on $\eta$.
It is obtained by integrating the original distribution on
the associated equivalent class. 
The conditional probability given $\tilde X$ is
\be\label{cond_BB}
P(\hat E_k|\tilde X)=\langle \tilde X|\hat E_k |\tilde X\rangle.
\ee
The probability distribution~(\ref{prob_BB}) and the
conditional probability~(\ref{cond_BB}) correspond to
the Beltrametti-Bugajski model. $\square$

Thus, any ontological theory that preserves the minimal
labelling for events of quantum mechanics necessarily
coincides with the Beltrametti-Bugajski model. In particular,
it is $\psi$-ontic~\cite{spekkens2}, a $\psi$-ontic theory being
an ontological theory of quantum mechanics that associates
two different quantum states with non-overlapping probability
distributions. In other words, the ontic state contains
the full information on the quantum state, which therefore 
represents some element of reality. This property was inferred in 
Ref.~\cite{morris} by using the hypothesis of POVM non-contextuality, 
which will be discussed in the next section. 
The Beltrametti-Bugajski model is clearly $\psi$-ontic because of the 
delta shape of the distribution defined by Eq.~(\ref{state_BB}).
Conversely,
in a $\psi$-epistemic theory, only probability distributions
associated with orthogonal states have disjoint supports. Thus,
it is not possible to infer the quantum state by the knowledge 
of the ontic state. In such a theory the quantum state does not 
represent an element of reality, but it contains a mere statistical 
information about the actual ontic state.
An example of $\psi$-epistemic model
is the Kochen-Specker (KS) model for a qubit~\cite{KS}. The 
ontological space is the set of Bloch vectors $\vec v$.
If a quantum state is represented by a Bloch vector $\vec b$,
the associated probability distribution is, up to a normalization
constant,
\be
\rho(\vec v|\vec b)=\theta(\vec v\cdot\vec b)\vec v\cdot\vec b,
\ee
where $\theta(x)$ is the Heaviside function. The support of this 
probability distribution is a hemisphere. It is clear that two 
probability distributions $\rho(\vec v|\vec b)$ and 
$\rho(\vec v|\vec b')$ are not overlapping only if $\vec b$ and 
$\vec b'$ are anti-parallel, corresponding to orthogonal quantum
states. By labelling an event with a Bloch vector $\vec c$,
the conditional probability of $\vec c$ given the ontic state 
$\vec v$ is, in the KS model,
\be
P(\vec c|\vec v)=\theta(\vec c\cdot\vec v).
\ee
This probability does not have
the structure given by Eq.~(\ref{gleason}) implied by
the Gleason theorem for a Hilbert space dimension greater
than $2$.

There exist other models for qubit satisfying the 
conditions~(\ref{prop1}-\ref{prop5}) but inequivalent
to the Beltrametti-Bugajski model, such as the Bell's
model~\cite{Bell} and a one-dimensional model recently reported
in Refs.~\cite{monti2,monti3}.

\section{Non-contextuality for unsharp measurements}
\label{sec2}
The generalized concept of non-contextuality can be also applied to 
unsharp measurements, that is, positive operator valued measurements
(POVM). They can be physically implemented by means of a
projective measurement on the system and an ancilla. A POVM is
defined by a set of positive operators $\{\hat Q_k\}$ satisfying
the condition 
\be
\sum_k\hat Q_k=\hat{\mathbb{1}}.
\ee
Each event is associated with an operator $\hat Q_k$. 
An ontological theory is non-contextual for unsharp measurements if the 
conditional probability for an event $\hat Q_k$ given an ontic state 
$X$ does not depend on the whole set of positive operators 
$\{\hat Q_1,\hat Q_2,...\}$ and, more in general, on the
physical implementation of the measurement.
It is a trivial task to prove the following. \newline
{\bf Lemma 1.} The Beltrametti-Bugajski theory is POVM non-contextual. 

Proof:
A POVM measurement is made on a system $A$ by
performing a measurement of the projectors $\hat E_k$ on 
$A$ and an ancilla $B$. Let $A$ be in a pure state. The overall
state is $\hat\rho=\hat\rho_A\hat\rho_B$, where
$\hat\rho_A=|\psi_A\rangle\langle\psi_A|$.

The positive operators $\hat Q_k$ are given by the equation
\be\label{effect}
\hat Q_k=\mathrm{Tr}[\hat\rho_B\hat E_k].
\ee
The probability of event $\hat Q_k$ is equal to the
probability of event $\hat E_k$, that is,
\be
p(\hat Q_k)\equiv\mathrm{Tr}_A[\hat Q_k\hat\rho_A]=
\mathrm{Tr}_{AB}[\hat E_k\hat\rho_A\hat\rho_B]\equiv \tilde p(\hat E_k)
\ee
At the ontological level, the probability of $\hat Q_k$
is given by
\be\label{sys_anci}
\begin{array}{c}
p(\hat Q_k)=\tilde p(\hat E_k)=  \\
\int P(\hat E_k|X_A,X_B)\delta(X_A-\psi_A)\rho_B(X_B)d X_A d X_B,
\end{array}
\ee
where the probability distribution $\rho_B$ satisfies the relation
\be\label{ancilla}
\hat\rho_B=\int |X\rangle\langle X|\rho_B(X) d X.
\ee
Note that this equation, because of the preparation contextuality for mixed 
states~\cite{spekkens}, is not necessarily invertible and many probability
distributions can correspond to the same density operator $\hat\rho_B$.
For example, the quantum state $\frac{1}{2}|\uparrow\rangle\langle\uparrow|+
\frac{1}{2}|\downarrow\rangle\langle\downarrow|$ is associated with
any probability distribution of the form $\frac{1}{2}\delta(X_B-\psi_1)+
\frac{1}{2}\delta(X_B-\psi_2)$, where $|\psi_1\rangle$ and $|\psi_2\rangle$
are two generic orthogonal states.
Integrating in $X_B$, Eq.~(\ref{sys_anci}) becomes
by means of Eqs.~(\ref{cond_BB},\ref{ancilla}) 
\be
p(\hat Q_k)=\int\mathrm{Tr}[|X_A\rangle\langle X_A|\hat\rho_B\hat E_k]
\delta(X_A-\psi_A)dX_A.
\ee
Tracing away the system $B$ and using Eq.~(\ref{effect}), we obtain that 
the probability of $\hat Q_k$ is given by
\be
p(\hat Q_k)=\int P(\hat Q_k|X_A) \delta(X_A-\psi_A)dX_A,
\ee
where 
\be
P(\hat Q_k|X)=\mathrm{Tr}_A(\hat Q_k|X\rangle\langle X|),
\ee
and the lemma is proved. $\square$

Thus, at the ontological level the conditional probability of an event 
depends only on the associated positive operator. This labelling
does not have memory of the particular projectors $\hat E_k$ and
ancilla state $\rho_B(X)$ that were used. Thus, the 
Beltrametti-Bugajski model is POVM non-contextual.
The proved theorem and lemma imply the following. \newline
{\bf Corollary:} The projective measurement non-contextuality implies
POVM non-contextuality for Hilbert dimensions greater than $2$. 

If we require the POVM non-contextuality for a qubit as an
additional hypothesis, it is easy to show, by using the 
generalized Gleason theorem in Ref.~\cite{busch}, that
the statement of the theorem proved in the previous section
holds in any dimension.
Indeed the models by Kochen and Specker, Bell and the recent one 
reported in Refs.~\cite{monti2,monti3} are all contextual
for unsharp measurements. Apart from the Bell model,
this is also a direct consequence of the fact that non-contextuality
for POVM implies $\psi$-onticity, as proved in Ref.~\cite{morris}.
Indeed neither the Kochen-Specker model or that in 
Refs.~\cite{monti2,monti3} are $\psi$-ontic.

\section{Extended quantum theory and macroscopic realism}
\label{sec3}
\subsection{Extended quantum theory}
In the Copenhagen interpretation, the quantum
state is not anything more than a mathematical tool for evaluating
probabilities. It merely represents the information about the 
preparation procedure of systems. Conversely, an ontological 
interpretation is designed to provide a realistic description of
systems through well-defined classical variables.
In this section we take a step
back and try a compromise between the two interpretations.
The technical results do not differ from those in Sec.~\ref{sec1}, 
but their interpretation is different and makes our results 
more closely related to recent findings~\cite{colbeck0}.

Instead of replacing the quantum state with ontic states, we can 
preserve it as a mathematical tool for describing preparation 
procedures and enrich the description by some additional information, 
provided by classical variables. It is important to stress that 
$|\psi\rangle$ is not supposed to describe some real entity, such as 
in de Broglie-Bohm mechanics and Beltrametti-Bugajski model. In this 
framework, an ensemble of systems, identically prepared in a pure state 
$|\psi\rangle$ within the context $\eta$, is described by the preparation 
protocol $\{|\psi\rangle,\eta\}$ and a probability distribution, 
$\rho(Y|\psi,\eta)$, of a classical variable $Y$, that is,
\be\label{preps}
\text{preparation}\rightarrow\{|\psi\rangle,\eta,\rho(Y|\psi,\eta)\}.
\ee
The probability of an outcome $\hat E_k$ is conditioned by the 
classical variable $Y$ and depends also on the overall preparation
procedure $\{|\psi\rangle,\eta\}$. Additionally,
for the moment we assume that it also depends on the measurement
context, specified by a parameter, $\tau$. We indicate
this conditional probability with $P(\hat E_k|Y,\tau,\psi,\eta)$.
Thus,
\be\label{events}
\text{event probability}\rightarrow P(\hat E_k|Y,\tau,\psi,\eta).
\ee

The model is equivalent to quantum theory if equation
\be\label{equiv1}
\int dY P(\hat E_k|Y,\tau,\psi,\eta)\rho(Y|\psi,\eta)=\langle\psi|\hat E_k
|\psi\rangle
\ee
is satisfied. The overall statements~(\ref{preps}-\ref{equiv1}) define 
an extended quantum theory, where quantum information is supplemented 
by classical information. The extension is said {\it non-trivial}
if the conditional probability $P(\hat E_k|Y,\tau,\psi,\eta)$
is not constant on the support of $\rho(Y|\psi,\eta)$. It provides
a fully ontological description if $P(\hat E_k|Y,\tau,\psi,\eta)$
does not depend on $\{\psi,\eta\}$. The Beltrametti-Bugajski
model is an example of trivial extension where the classical
variables replace completely $\psi$. In practise, $\psi$ is
reinterpreted as a physical field, but no further information is
introduced that provides a more accurate predictions of outcomes.
Just as quantum theory is formally identical to the 
Beltrametti-Bugajski model and differs only in the interpretation,
any trivial extension of quantum theory is formally identical
to the Beltrametti-Bugajski model.
In general, an extended quantum theory is formally identical to 
a $\psi$-ontic theory where $(\psi,Y)$ are the ontological variables, 
with the only difference that the quantum state $|\psi\rangle$ 
is interpreted as a container of information about the preparation 
procedure. In other words, an extended theory is a mixture of an operational
and ontological description. 

The Kochen-Specker theorem establishes that there is no deterministic 
ontological theory that is non-contextual for measurements. However, this 
result does not rule out the possibility of supplementing the quantum 
state with some amount of classical information and simultaneously 
preserving the minimal labelling of events through projective
operators (without $\tau$). So in principle, we 
could have a non-trivial extended quantum theory defined by 
statements~(\ref{preps}-\ref{equiv1}) and satisfying
the condition
\be\label{noncont}
P(\hat E_k|Y,\tau,\psi,\eta)=P(\hat E_k|Y,\psi,\eta).
\ee
With a slight abuse of terminology, we call this condition
{\it non-contextuality} for measurements. 
This expanded definition is justified by the following. \newline
{\bf Lemma 2.} If there is a (non-trivial) extended quantum
theory that is non-contextual for measurements, then there is a 
(non-trivial) ontological theory with the same
property. The converse is also true.

By non-trivial ontological theory we mean a theory that
is essentially different from the Beltrametti-Bugajski model.
The proof of this lemma is very simple. Indeed, every
extended theory generates an ontological theory with $(\psi,Y)$ 
(and possibly the preparation context $\eta$) as ontological variables. 
Furthermore, if the extended theory is non-trivial, also the generated 
ontological theory is non-trivial. It is obvious that the property of 
non-contextuality is preserved in this change of interpretation.
The converse is also true, since a (non-trivial) ontological theory 
is a (non-trivial) extended quantum theory where the conditional 
probabilities for events do not depend on $\psi$.

By using the results in Sec.~\ref{sec1}, it is easy to prove
the following.
\newline
{\bf Theorem 2.} Quantum theory cannot be extended in a non-trivial
way by preserving the minimal labelling for events of quantum
mechanics. In other words, there is not a non-trivial extended
quantum theory that is non-contextual for measurements.

In spite of the different interpretation, this theorem is
a rephrasing of the theorem~1 proved in Sec.~\ref{sec1}. Indeed,
as previously said, an extended theory is formally identical
to an ontological theory, where $(\psi,Y)$ are the ontological
variables. We have proved in Sec.~\ref{sec1} that a such theory 
is essentially equivalent to the Beltrametti-Bugajski model under 
the hypothesis of non-contextuality for measurements, that is,
$Y$ does not introduce any improvement in the prediction of
outcomes. Indeed, as stated by lemma~2, if there was a non-contextual
and non-trivial extended theory, then there would be a non-contextual 
ontological theory essentially different from the Beltrametti-Bugajski
model.

A similar theorem was proved in Ref.~\cite{colbeck0}, where the
authors used a hypothesis of statistical independence between
the setting parameters of a measurement and the variables that 
do not lie in the future light cone of those parameters 
({\it free choice hypothesis}). By means of this hypothesis 
they derived a relation similar to Eq.~(\ref{noncont}),
which directly implies that quantum mechanics cannot be extended. 
This was initially
accomplished in the case of maximally entangled states and
subsequently generalized by means of the hypothesis (called QMb) 
that every quantum process is unitary. Our proof has the advantage 
of simplicity, resting upon the well-established and
powerful Gleason theorem~\cite{G}. Furthermore it does not 
need any additional hypothesis on the dynamics, such as QMb.

\subsection{Macroscopic realism}

It is interesting to note that macroscopic realism requires
that some amount of classical information has to be supplied
to the quantum state, provided that the quantum
alternatives have reached some level of "macroscopicness". In order 
to be complete, a theory should contain the description of
this classical information in its formalism. Suppose for
example that a microscopic quantum system, {\it A}, is in the 
superposition 
$|1\rangle+|-1\rangle$ and interacts with a macroscopic
device, {\it D}, which is initially in the state $|\varphi_0\rangle$. 
After the interaction the overall quantum system evolves towards
the entangled state
$$
|\Psi\rangle=|1\rangle|\varphi_1\rangle+|-1\rangle|\varphi_{-1}\rangle,
$$
where $|\varphi_{\pm1}\rangle$ are two macroscopically distinct 
states, for example corresponding to different positions of a
pointer. The state $|\Psi\rangle$ contains the information
about the initial preparation of the system {\it A}+{\it D}
and its subsequent evolution. Macroscopic realism
imposes that this information has to be supplemented
with some classical information indicating the actual
macroscopic state of the device. 
The classical information can be stored
in a binary variable, $n=\pm1$, where $\pm1$ correspond
to the macroscopic states $|\varphi_{\pm1}\rangle$.
In this extended description, the overall information on
the system {\it A}+{\it D} is given by the pair
$$
(|\Psi\rangle,n)
$$
and, additionally, other parameters describing the context.
In general, the outcome of every external observation
performed on  {\it A}+{\it D} has a probability
that depends on both $|\Psi\rangle$ and $n$.
Indeed, the probability of finding the device
in the state $|\varphi_\pm\rangle$ has to be equal to $1$
or $0$, provided that $n=\pm1$ or $n=\mp1$. In other
words, the outcome is completely determined for the measurement 
of the "pointer" state and the extension cannot be trivial.
Thus, macroscopic realism imposes that quantum theory has
to be extended in a non-trivial way with some amount of classical 
information and this cannot be accomplished without giving up 
non-contextuality for measurement. In this inference we have implicitly 
used the additional hypothesis of unitarity for evolutions, as discussed 
in the following paragraph.

It is worthwhile to note that in practise it is not possible 
to perform every kind of measurements on a macroscopic system 
and the events that can be actually observed are not in general 
affected by the replacement of the superposition of
$|1\rangle|\varphi_1\rangle$ and $|-1\rangle|\varphi_{-1}\rangle$
with their mixture. This property is called {\it decoherence}
and provides a justification to the quantum state reduction,
which is one of the postulates of quantum theory.
It implies a loss of information, for which the states
$(|\Psi\rangle,\pm1)$  are indistinguishable from the
states $(|\pm1\rangle|\varphi_{\pm1}\rangle,\pm1)$. 
The reduction postulate is particularly relevant in this context
because the quantum state collapse into macroscopic distinct
states would make unnecessary the addition of classical
information and would remove the contradiction between
non-contextuality for measurements and macroscopic realism.
In our opinion, the quantum state collapse is not more fundamental 
than the loss of information in statistical mechanics and
in fact the detailed full information is provided by the state 
$(|\Psi\rangle,\pm1)$. Indeed, in principle nothing
forbids one to perform a measurement that is able to 
distinguish a superposition of states from their mixture.
This is particularly true if the device is
mesoscopic. Promoting the decoherence to the
rank of fundamental principle would raise the issue
of deciding what is the level of "macroscopicness" and
decoherence above which a superposition is
replaced by a mixture. 

We conclude this section by noting that an argument that
uses decoherence as a fundamental principle for explaining
macroscopic realism implicitly seems to require a weakening of 
causality principle. Indeed, according to this explanation, one 
system is in a defined macroscopic system because never in the future 
there will be an emerging property revealing interference between
macroscopically distinct states. This point is made more explicit
in the {\it consistent histories} approach to quantum mechanics~\cite{griffiths}, 
where the condition for consistency involves the whole temporal history.
The weakening of causality principle was also suggested in recent 
papers~\cite{monti2,monti3,monti4} as a solution of the problem
of the exponential growth of resources that are required for 
specifying an ontic state. However, the challenge of undermining
the causality principle is beyond the purpose of this paper.

\section{Conclusions}

We have proved that the Beltrametti-Bugajski model is in practise
the only model that is non-contextual for projective measurements
and Hilbert space dimension $N$ greater than $2$. Equivalently,
it is not possible to extend non-trivially quantum theory without
giving up non-contextuality for measurements.
As a corollary,
we have shown that POVM non-contextuality for $N>2$ is implied by sharp
measurement non-contextuality. The theorem can be generalized to
a qubit by employing the additional hypothesis of POVM 
non-contextuality for $N=2$. 

The Beltrametti-Bugajski model is the simplest example of ontological 
theory of a measurement process. However it cannot be considered a completely 
realistic theory.
In particular, since it still needs to invoke a measurement 
made by something external to the system, in fact does not solve 
the measurement problem without retaining the postulate of
wave-function collapse required in the standard intepretation.
An exhaustive realistic model should at least satisfy a criterion of 
macroscopic realism, attributing for example a sufficiently well
defined value to the position of macroscopic objects that does not
depend on a possible external observation. This is the case of the 
de Broglie-Bohm mechanics, where the wave-function is
supplied by additional variables describing the positions of
all the particles compounding a system. The Beltrametti-Bugajski
model fails to satisfy the criterion of macroscopic realism.
Suppose for example that the position of an object is in 
the superposition of two macroscopically separated values. 
The Beltrametti-Bugajski model, being in fact a rephrasing of the quantum 
mechanics, does not give any description of the actual position 
of that object, unless a breaking of unitarity is invoked through
the quantum state collapse.
Indeed, any non-trivial extension of the theory acted to introduce 
macroscopic realism without breaking unitarity would automatically 
make the theory measurement contextual. Thus, the only way to solve
the measurement problem in the framework of an ontological
theory is relaxing the hypothesis of measurement non-contextuality 
in the generalized sense introduced in Ref.~\cite{spekkens}.

We conclude noting that macroscopic realism implies a certain
degree of outcome determinism, as discussed in Sec.~\ref{sec3}.
Given a macroscopic system, there should be always a set of macroscopic 
states such that the system 
is in one of them. The result of a possible measurement, acted to
know that state, would be completely determined 
by the hidden variable state of the system. It is interesting to observe
that the criterion of macroscopic realism is rarely taken in consideration
in the study of ontological models.
We have shown that it imposes some constraints and is very useful 
to deduce general properties that an ontological theory must satisfy,
such as measurement contextuality.

\section*{Acknowledgments}
\nonumber
A. M. acknowledges useful discussions with Robert W. Spekkens,
Brian Morris and Roger Colbeck.
This work was partially supported by the National Natural Science
Foundation of China under Grant No.10775175.
Research at Perimeter Institute for Theoretical Physics is
supported in part by the Government of Canada through NSERC
and by the Province of Ontario through MRI.

\end{document}